\documentstyle[psfig]{neu}
\begin{document}

\title{%
RXTE/ASM OBSERVATIONS\\ OF PSR~B1259$-$63}

\author{Victoria M. KASPI and Ronald A. REMILLARD\\
{\it    Department of Physics and Center for Space Research,
Massachusetts Institute of Technology, Cambridge, MA 02139, USA\\
vicky@space.mit.edu, rr@space.mit.edu}
}

\maketitle

\section*{Abstract}

{\it RXTE}/ASM observations of PSR~B1259$-$63 near the 1997 periastron
passage find no emission brighter than $\sim$10 times that
detected by {\it ASCA} during the 1994 periastron encounter,
demonstrating that the pulsar is unlikely to have undergone even brief
episodes of accretion.

\section{Introduction, Observations, and Conclusions}

The 48~ms radio pulsar PSR~B1259$-$63 is in a highly eccentric 3.4~yr
orbit with a Be star (Johnston et al. 1992).  Near periastron, for a
sufficiently strong Be star wind, PSR~B1259$-$63 could possibly make a transition
to an accretion regime, exhibiting X-ray pulsations of
luminosity $>$10$^{35}$~erg~s$^{-1}$.  {\it ASCA} X-ray observations at
three epochs around the 1994 periastron passage (Kaspi et al. 1994;
Hirayama 1996) found the source to be unpulsed, with modest luminosity
$\sim$10$^{34}$~erg~s$^{-1}$, for a distance of 2~kpc.  These
properties imply that shock emission produces the
observed X-rays, rather than accretion (Tavani \& Arons 1997).
However radio timing observations suggested sudden, brief accretion events near
periastron that could not be
ruled out by the X-ray observations (Manchester et al. 1995).  Indeed the pulsar's anomalously
low period suggests it may have undergone occasional accretion
episodes in the past.  {\it ASCA} observations during the 1997
periastron passage were impossible due to solar constraints.

The ASM on {\it RXTE} has been observing bright celestial X-ray
sources regularly since early 1996. The instrument consists of
three ``scanning shadow cameras,'' each containing a
position-sensitive proportional counter that is mounted below a
wide-field collimator, covered by a coded mask.  The
instrument provides roughly 5 celestial scans per day, with diminished
exposure in directions toward the Sun.  Further information on the ASM
is given by Levine et al. (1996).  While PSR~B1259$-$63 is not
routinely monitored for the ASM source-history
database, we extracted an X-ray light curve as an archival analysis
project using standard techniques. The derived light curve
(Figure~1a) covers the time interval of MJD 50178--50762 (1996 Apr 5
to 1997 Nov 10).  The overall mean intensity of PSR B1259$-$63 during
this time interval was $(0.5 \pm 2.4)\times
10^{-11}$~erg~s$^{-1}$~cm$^{-2}$ at 2--10 keV. This result is given
with consideration of both the systematic bias and uncertainty in the
ASM light curves for faint, yet fairly isolated X-ray sources.
Figure~1b shows $3\sigma$ upper limits obtained in the periastron
vicinity in 2-day averages.  It shows that in the 2 weeks following
periastron, the source was not observed to be brighter than $\sim$10
times the 1994 periastron luminosity.

The $3\sigma$ upper limits to the X-ray emission from PSR~B1259$-$63
from the ASM thus argue that the pulsar did not undergo even brief
episodes of accretion during the 1997 periastron passage.  This is
consistent with the shock emission model, as well as with new
conclusions based on timing that rule out the sudden ``spin-ups'' that
were claimed previously (Wex et al. 1998).

\begin{figure}[t]
\centerline{
\psfig{figure=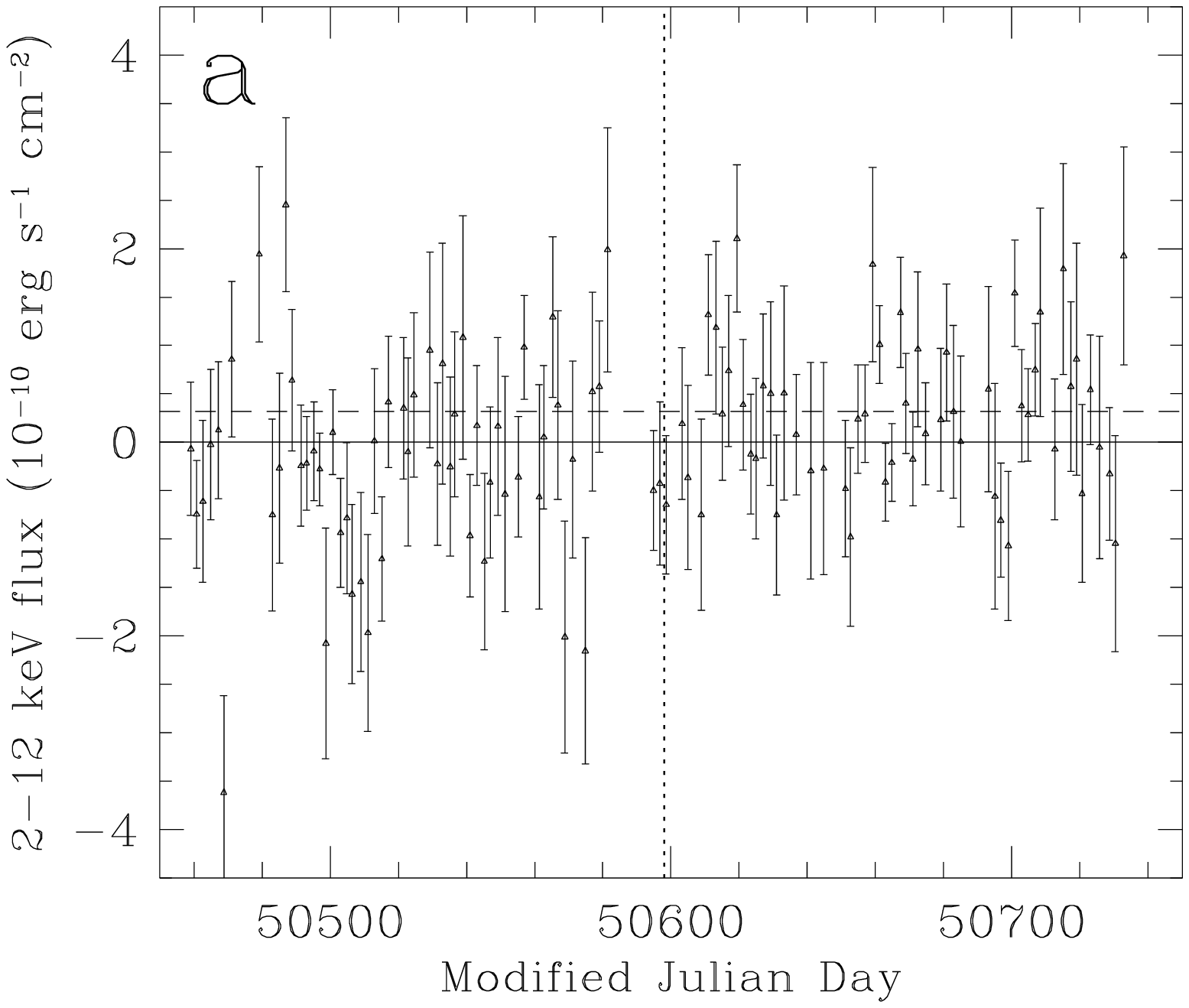,height=5.5cm}
\psfig{figure=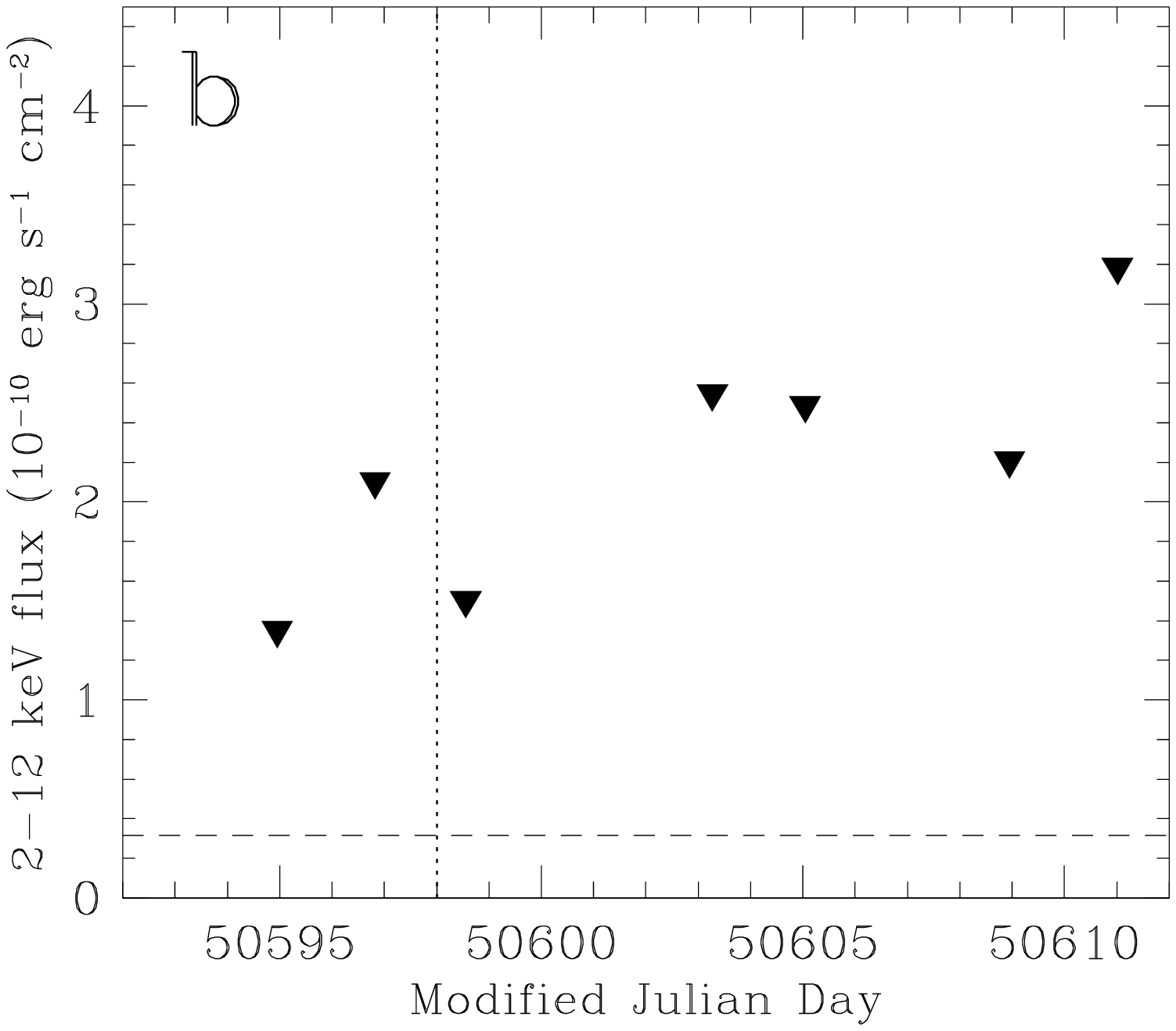,height=5.5cm}
}
\caption{(a) ASM light curve for PSR~B1259$-$63. (b) 3$\sigma$
upper limits to the ASM flux near periastron from (a).
The horizontal dashed line represents the brightest {\it
ASCA} 1994 periastron detection (Kaspi et al. 1994), and
the vertical dotted line indicates periastron.}
\end{figure}

\section{References}

\re
Hirayama, M. 1996, PhD Thesis, University of Tokyo
\re
Kaspi, V. M. et al. 1995, ApJ, 453, 424
\re 
Levine, A. M. et al. 1996, ApJ, 469, L33
\re
Manchester, R. N. et al. 1995, ApJ, 445, L137
\re
Tavani, M., Arons, J. 1997, ApJ, 477, 439
\re
Wex, N. et al. 1998, MNRAS, in press

\end{document}